\documentclass{jpp}

\usepackage{graphicx,bm}
\usepackage{natbib}
\newcommand{\bec}[1]{\mbox{\boldmath $ #1$}}
\newcommand{\EQ}{\begin{equation}}
\newcommand{\EN}{\end{equation}}
\newcommand{\EQA}{\begin{eqnarray}}
\newcommand{\ENA}{\end{eqnarray}}

\newcommand{\meanrho}{\overline{\rho}}
\newcommand{\meanE}{\overline{E}}

{}
{}
{}

{}
{}
{}
{}
{}
{}
{}
{}
{}
{}
{}
{}
{}
{}
{}
{}
{}
{}
{}
{}
\newcommand{\meanUU}{\overline{\mbox{\boldmath $U$}}{}}{}

{}
{}
{}

\newcommand{\meanp}{\overline{P}}
\newcommand{\means}{\overline{S}}
\newcommand{\meanT}{\overline{T}}
\newcommand{\meanW}{\overline{W}}
\newcommand{\meanQ}{\overline{Q}}

{}

{}
{}

\newcommand{\nab}{\mbox{\boldmath $\nabla$} {}}

\def\urms{u_{\rm rms}}

\def\half{{\textstyle{1\over2}}}

\def\onethird{{\textstyle{1\over3}}}

\title[Turbulent fluxes of entropy and internal energy]{Turbulent fluxes of entropy and internal energy in temperature stratified flows}

\author[I. Rogachevskii and N. Kleeorin]%
{I.\ns R\ls O\ls G\ls A\ls C\ls H\ls E\ls V\ls S\ls K\ls I\ls I
  \thanks{Email address for correspondence: gary@bgu.ac.il}
\and
N.\ns K\ls L\ls E\ls E\ls O\ls R\ls I\ls N}

\affiliation{
Department of Mechanical Engineering, Ben-Gurion University of
the Negev, P. O. Box 653, 84105 Beer-Sheva, Israel\\[\affilskip]
Nordita, KTH Royal Institute of Technology
and Stockholm University, Roslagstullsbacken 23,
10691 Stockholm, Sweden
}

\pubyear{2010}
\volume{650}
\pagerange{119--126}

\date{?; revised ?; accepted ?. }
\begin{document}

\maketitle

\begin{abstract}
We derive equations for the mean entropy and the mean internal energy
in the low-Mach-number temperature stratified turbulence (i.e., for
turbulent convection or stably stratified turbulence),
and show that turbulent flux of entropy is given by ${\bm F}_s=\meanrho \, \overline{{\bm u} s}$, where  $\meanrho$ is the mean fluid density, $s$ are fluctuations of entropy
and overbars denote averaging over an ensemble of turbulent velocity field, ${\bm u}$. We
demonstrate that the turbulent flux of entropy is different from the turbulent
convective flux, ${\bm F}_c=\meanT \, \meanrho \,
\overline{{\bm u} s}$, of the fluid internal energy,
where $\meanT$ is the mean fluid temperature.
This turbulent convective flux is well-known in the astrophysical and geophysical literature,
and it cannot be used as a turbulent flux in the equation for the mean entropy.
This result is exact for low-Mach-number temperature stratified turbulence
and is independent of the model used.
We also derive equations for the velocity-entropy correlation, $\overline{{\bm u} s}$, in the limits of small and large P\'{e}clet numbers, using the quasi-linear
approach and the spectral $\tau$ approximation, respectively.
This study is important in view of different applications to the astrophysical and geophysical temperature stratified turbulence.
\end{abstract}

\section{Introduction}

Temperature stratified turbulence (e.g., turbulent convection or stably
stratified turbulence) plays a crucial role in astrophysics
\citep{SSZ78,P80,ZRS83,ZRS90,RSS88,CC07} and
geophysics \citep{MY75,Z91,C09,ZEKR08,ZEKR13}. The large-scale properties of the
temperature stratified turbulence are determined in the framework
of the mean-field approach in which all quantities are decomposed into the
mean and fluctuating parts, where the fluctuating parts
have zero mean values and overbars denote averaging
over an ensemble of turbulent velocity fields.

In the astrophysical and geophysical literature
on the low-Mach-number temperature stratified turbulence
two different formulae for the turbulent flux of
entropy are used. The first formula coincides with the turbulent
convective flux of internal energy,
${\bm F}_c=\meanT \, \meanrho \,\, \overline{{\bm u} s}$, so that
the equation for the mean entropy is \citep{KM00,BMT04,MB08,JKM09,JK09,KMB12}:
\begin{eqnarray}
\meanrho \left({\partial \means \over \partial t} + (\meanUU \cdot \nab)\means \right) + {1 \over \meanT} \nab \cdot (\meanT \, \meanrho \, \, \overline{{\bm u} s}) =
{1 \over \meanT} \left[\nab\cdot(K \nab \meanT)
+ \overline{J}\,\right] ,
\label{A1}
\end{eqnarray}
where $\meanrho$, $\meanT$, $\means$ and $\meanUU$ are the mean fluid density, temperature,
specific entropy and mean velocity, respectively; ${\bm u}$ and $s$ are the fluctuations
of fluid velocity and entropy, respectively;
$\overline{J}$ is the mean source and/or sink of the entropy (that also includes the
viscous heating) and $K$ is the coefficient of the molecular heat conductivity.
The last term on the left-hand side of Eq.~(\ref{A1}) corresponds to the
turbulent flux of entropy.

The other form of the turbulent flux of
entropy is ${\bm F}_s=\meanrho \, \overline{{\bm u} s}$,
and the equation for the mean entropy is \citep{BR95,GR96b,GR96a}:
\begin{eqnarray}
\meanrho \left({\partial \means \over \partial t} + (\meanUU \cdot \nab)\means \right) + \nab \cdot (\meanrho \, \overline{{\bm u} s}) =
{1 \over \meanT} \left[\nab\cdot(K \nab \meanT)
+ \overline{J}\,\right].
\label{A2}
\end{eqnarray}
Equations~(\ref{A1}) and~(\ref{A2}) are essentially different. In particular,
the last term on the left-hand sides of Eq.~(\ref{A1}) and~(\ref{A2}) are different.

The goal of the present paper is to derive equations for the mean entropy and the mean internal energy which yield formulae for the turbulent flux of
entropy and the turbulent flux of internal energy, and to clarify which equation
for the mean entropy [Eq.~(\ref{A1}) or~(\ref{A2})] used in the temperature
stratified turbulence, is correct.
When the fluid temperature profile is not uniform, the above question is crucial.

\section{Turbulent convective flux of mean internal energy and turbulent flux of mean entropy}

In this Section we will derive equations for the mean entropy and the mean internal energy.
We consider low-Mach-number temperature stratified fluid flows.

\subsection{Governing equations}

The budget equation for the instantaneous internal
energy density $E= c_{\rm v} T$ is \citep{LL59}:
\begin{eqnarray}
{\partial (\rho \, E )\over \partial t} +
\nab \cdot \left(\rho \,{\bm U} W - {\bm U} P - K \nab T \right)
= Q,
\label{AD1}
\end{eqnarray}
where ${\bm U}$ is the instantaneous velocity determined by the Navier-Stokes equation
for fluid motion,
$\rho$, $T$ and $P$ are the instantaneous density, temperature and pressure, respectively,
which satisfy the equation of state for a perfect gas,
$K$ is the coefficient of the molecular heat conductivity,
$W=c_{\rm p} T = c_{\rm v} T + P/\rho= E + P/\rho$ is the instantaneous enthalpy,
where $c_{\rm v}$ and $c_{\rm p}$ are the specific heats
at the constant volume and pressure, and
$Q= - P \nab~\cdot~{\bm U} + \hat{\sigma}_{ij}({\bm U}) \nabla_j U_i$, where
$\hat{\sigma}({\bm U}) = 2 \nu \rho \hat{\cal S}({\bm U})$, $\hat{\cal S}({\bm U})
={\cal S}_{ij}=\half(U_{i,j}+U_{j,i})
-\onethird\delta_{ij}\nab\cdot{\bm U}$,
$\nu$ is the kinematic viscosity, and $\delta_{ij}$ is the Kronecker tenzor.
The instantaneous density $\rho$ is determined by the continuity equation:
\begin{eqnarray}
{\partial \rho\over \partial t} + \nab \cdot (\rho \, {\bm U}) =0.
\label{AD6}
\end{eqnarray}
The budget equation for the instantaneous kinetic energy
density $\half\rho {\bm U}^2$ is \citep{LL59}:
\begin{eqnarray}
{\partial \over \partial t} \left(\half\rho {\bm U}^2\right) +
\nab \cdot \left[{\bm U} \left(\half\rho {\bm U}^2+ P\right) - {\bm U} \hat{\sigma}({\bm U}) \right]
= -Q .
\label{AD9}
\end{eqnarray}
The sum of Eqs.~(\ref{AD1}) and~(\ref{AD9}) yields the conservation law for the instantaneous total (kinetic plus internal) energy densities $\half\rho {\bm U}^2 + \rho \, E$ \citep{LL59}:
\begin{eqnarray}
{\partial  \over \partial t} \left(\half\rho {\bm U}^2 + \rho \, E\right) +
\nab \cdot \left[{\bm U}\left(\half\rho {\bm U}^2 + \rho W\right) - {\bm U} \hat{\sigma}({\bm U})
- K \nab T \right] =0 .
\label{AD12}
\end{eqnarray}

\subsection{Turbulent convective flux and equation for mean internal energy}

Averaging Eq.~(\ref{AD1}) over the ensemble we obtain
the budget equation for the mean internal energy density
$\meanE= c_{\rm v} \meanT$:
\begin{eqnarray}
{\partial (\meanrho \, \meanE) \over \partial t} +
\nab \cdot \left(\meanrho \, \meanUU \, \meanE + \meanrho \, \overline{{\bm u} w}
- \overline{{\bm u} p}- K \nab \meanT\right) = \meanQ,
\label{AD2}
\end{eqnarray}
where ${\bm u}$ are the velocity fluctuations, $w= \meanT \, s + p/\meanrho$ are the enthalpy fluctuations, $s$ and $p$ are the entropy and pressure fluctuations, respectively, and
$\meanQ= - \meanp \nab \cdot \overline{\bm U} - \overline{p \nab \cdot {\bm u}} + \hat{\sigma}_{ij}(\overline{\bm U}) \nabla_j \overline{U}_i + \overline{\hat{\sigma}_{ij}({\bm u}) \nabla_j u_i}$.
In the derivation of Eq.~(\ref{AD2}) we used the identity:
$\meanW = \meanE + \meanp/\meanrho$, and since
we consider a low-Mach-number turbulent flow, we took into account that
$|\overline{{\bm u} \rho}|/ |\meanrho|  \ll
|\overline{{\bm u} s}|/|\means|$ \citep{CHA03}.
The turbulent flux of enthalpy is
\begin{eqnarray}
\overline{{\bm u} w} = \meanT \, \overline{{\bm u} s}
+ {\overline{{\bm u} p} \over \meanrho}.
\label{AD3}
\end{eqnarray}
Since $\meanW = \meanE + \meanp/\meanrho$ we obtain that $\meanrho \, \meanUU \, \meanE =
\meanUU \, (\meanrho \meanW - \meanp)$.
Substituting the latter equation and Eq.~(\ref{AD3}) into Eq.~(\ref{AD2}), we obtain:
\begin{eqnarray}
{\partial \meanrho \, \meanE \over \partial t} +
\nab \cdot \left(\meanrho \, \meanUU \, \meanE + \meanT \, \meanrho \,  \, \overline{{\bm u} s}
- K \nab \meanT\right)= \meanQ .
\label{AD4}
\end{eqnarray}
The mean fluid density $\meanrho$ is determined by the continuity equation:
\begin{eqnarray}
{\partial \meanrho\over \partial t} + \nab \cdot (\meanrho \, \meanUU) =0.
\label{AD15}
\end{eqnarray}
Equations~(\ref{AD4}) and~(\ref{AD15}) yield the following equation
for the evolution of the mean internal energy:
\begin{eqnarray}
\meanrho \left({\partial \meanE \over \partial t} + (\meanUU \cdot \nab)\meanE \right)
+ \nab \cdot \left(\meanT \, \meanrho \,  \, \overline{{\bm u} s}
- K \nab \meanT\right)= \meanQ ,
\label{ND4}
\end{eqnarray}
where the turbulent convective flux of the mean internal energy is:
\begin{eqnarray}
{\bm F}_c=\meanT \, \meanrho \,\, \overline{{\bm u} s} .
\label{ND5}
\end{eqnarray}

\subsection{Equation for the sum of mean and turbulent kinetic energies and
conservation law for total mean energy}

Averaging Eq.~(\ref{AD9}) for the instantaneous kinetic energy
density $\half\rho {\bm U}^2$ we obtain
an equation for the sum of the mean and turbulent kinetic energies $\half\meanrho \overline{\bm U}^2 + \half\meanrho \overline{{\bm u}^2}$:
\begin{eqnarray}
{\partial \over \partial t} \left(\half\meanrho \overline{\bm U}^2 + \half\meanrho \overline{{\bm u}^2} \right)   +
\nab \cdot \left[\overline{\bm U} \left(\half\meanrho \overline{\bm U}^2 + \meanp\right)
- \overline{\bm U} \hat{\sigma}(\overline{\bm U}) + \overline{{\bm u} \left(\half \rho {\bm u}^2 + p\right)}  - \overline{{\bm u} \hat{\sigma}({\bm u})} \right]
= -\overline{Q} .
\nonumber\\
\label{ND9}
\end{eqnarray}
The sum of Eqs.~(\ref{AD4}) and~(\ref{ND9}) yields the conservation
law for the total mean energy $E_{\rm tot}=\half\meanrho \overline{\bm U}^2 + \half\meanrho \overline{{\bm u}^2} + \meanrho \, \meanE$:
\begin{eqnarray}
&& {\partial \over \partial t} \left(\half\meanrho \overline{\bm U}^2 + \half\meanrho \overline{{\bm u}^2} + \meanrho \, \meanE\right)   +
\nab \cdot \biggl[\overline{\bm U} \left(\half\meanrho \overline{\bm U}^2 + \meanp + \meanrho \, \meanE\right) - \overline{\bm U} \hat{\sigma}(\overline{\bm U})
\nonumber\\
\quad &&- \overline{{\bm u} \hat{\sigma}({\bm u})} + \overline{{\bm u} \left(\half \rho {\bm u}^2 + p\right)} + \meanT \, \meanrho \,  \, \overline{{\bm u} s}
- K \nab \meanT \biggr]
= 0 .
\label{ND10}
\end{eqnarray}
This equation contains the turbulent convective flux ${\bm F}_c=\meanT \, \meanrho \, \overline{{\bm u} s}$.
The conservation law~(\ref{ND10}) for the total mean energy $E_{\rm tot}$ can be rewritten in terms of the mean, $\overline{\bm U} \, \, \meanW$, and turbulent, $\overline{{\bm u} w}$, fluxes of enthalpy [see Eq.~(\ref{AD3})]:
\begin{eqnarray}
&& {\partial \over \partial t} \left(\half\meanrho \overline{\bm U}^2 + \half\meanrho \overline{{\bm u}^2} + \meanrho \, \meanE\right)   +
\nab \cdot \biggl[\overline{\bm U} \left(\half\meanrho \overline{\bm U}^2 + \meanrho \, \meanW\right) - \overline{\bm U} \hat{\sigma}(\overline{\bm U})
\nonumber\\
\quad &&- \overline{{\bm u} \hat{\sigma}({\bm u})} + \overline{{\bm u} \left(\half \rho {\bm u}^2\right)} + \meanrho \, \overline{{\bm u} w}
- K \nab \meanT \biggr]
= 0 .
\label{ND11}
\end{eqnarray}

\subsection{Equation for mean entropy}

The evolution equation for the instantaneous entropy
$S=c_{\rm v} \, \ln(P \rho^{-\gamma})$ is
\citep{LL59}:
\begin{eqnarray}
\left({\partial \over \partial t} + {\bm U} \cdot \nab\right) S
= {1 \over \rho T} \left[\nab\cdot(K \nab T) + J\right],
\label{D5}
\end{eqnarray}
where $\gamma=c_{\rm p}/c_{\rm v}$ is the ratio of specific heats and
$J$ is a source and/or sink of the entropy (that also includes the
viscous heating).
Multiplying the equation for the entropy~(\ref{D5}) by the fluid density $\rho$, and the continuity equation~(\ref{AD6}) by the fluid entropy $S$, and
add them, we obtain the following equation:
\begin{eqnarray}
{\partial (\rho \, S) \over \partial t} + \nab \cdot (\rho \, {\bm U} S) =
{1 \over T} \left[\nab\cdot(K \nab T) + J\right].
\label{D7}
\end{eqnarray}
The second term on the left-hand sides of Eqs.~(\ref{D5}) and~(\ref{D7}),
which contributes to the turbulent diffusion of the mean entropy,
does not contain the temperature field. This is a reason why
the turbulent flux of the mean entropy
for stratified turbulence cannot contain the mean temperature.

Averaging Eq.~(\ref{D7}) over the ensemble we obtain
the equation for the mean entropy, $\means$:
\begin{eqnarray}
{\partial (\meanrho \, \means) \over \partial t} + \nab \cdot
\left(\meanrho \, \meanUU \, \means + \meanrho \, \overline{{\bm u} s}\right) =
{1 \over \meanT} \left[\nab\cdot(K \nab \meanT)
+ \overline{J}\,\right] .
\label{D8}
\end{eqnarray}
In the derivation of Eq.~(\ref{D8}) we have taken into account that
for a low-Mach-number turbulent flow:
$|\overline{\rho \, s}| \ll |\meanrho| \, |\means|$,
$|\overline{\rho \, s}|/ |\meanrho|  \ll |\overline{{\bm u} s}|/\urms$
and $|\overline{{\bm u} \rho}|/ |\meanrho|  \ll
|\overline{{\bm u} s}|/|\means|$
\citep{CHA03}. To get the simplest form of the molecular diffusion term and the source term
on the right-hand side of Eq.~(\ref{D8}) we assumed that: (a) the temperature fluctuations, $\theta$, are much smaller than the mean fluid temperature, $\meanT$, i.e., $|\theta| \ll \meanT$; (b) in the framework of the mean-field theory there is a separation of scales, $\ell_0 \ll L_T$, where  $L_T$ is the characteristic scale of the mean temperature variation and $\ell_0$ is the integral scale of turbulence (the random velocity field); (c) the coefficient of the molecular heat conductivity $K$ is independent of the temperature fluctuations and (d) fluctuations of the source or sink of the entropy, $J'$, are independent
of the temperature fluctuations.
Equations~(\ref{AD15}) and~(\ref{D8}) yield the following equation
for the evolution of the mean entropy:
\begin{eqnarray}
\meanrho \left({\partial \means \over \partial t} + (\meanUU \cdot \nab)\means \right) + \nab \cdot (\meanrho \, \overline{{\bm u} s}) =
{1 \over \meanT} \left[\nab\cdot(K \nab \meanT)
+ \overline{J}\,\right],
\label{NA2}
\end{eqnarray}
which coincides with Eq.~(\ref{A2}), and the turbulent flux of the mean entropy
for stratified turbulence with non-uniform profiles
of the mean fluid temperature and density is:
\begin{eqnarray}
{\bm F}_s=\meanrho \,\overline{{\bm u} s} .
\label{NA3}
\end{eqnarray}
Other forms of the turbulent flux of entropy [see Eq.~(\ref{A1})] used
in the astrophysical and geophysical literature are incorrect. This is an exact statement
for a low-Mach-number temperature stratified turbulence
and is independent of the model.

\section{The velocity-entropy correlation}

Now let us determine the velocity-entropy correlation, $\overline{{\bm u} s}$.
For simplicity we consider turbulent flows with a zero mean velocity, $\meanUU=0$.
Subtracting Eq.~(\ref{D8}) from Eq.~(\ref{D7}) we obtain the equation for the entropy fluctuations:
\begin{eqnarray}
{\partial s \over \partial t} + {\cal N} - \chi \bec{\nabla}^2 s  = I ,
\label{C1}
\end{eqnarray}
where ${\cal N} = (\meanrho)^{-1} \bec\nabla {\bm \cdot} \, [\meanrho({\bm u} s - \overline{{\bm u} s})]$  is the nonlinear term, $\chi=K/c_{\rm p}\meanrho$ is the molecular diffusion coefficient of entropy and $I = - (\meanrho)^{-1}\nab \cdot (\meanrho  \means \, {\bm u}) = - ({\bm u} \cdot \nab)\means$ is the source term. In the derivation of Eq.~(\ref{C1})  we took into account the anelastic approximation, $[\nab \cdot (\meanrho  \, {\bm u})=0$], we
assumed that fluctuations of the source or sink of the entropy, $J'$, are very small,
and we also assumed that the molecular diffusion term can be simplified as $(\meanrho \meanT)^{-1}\nab \cdot [(K/c_{\rm p}) \nab(\meanT s)] \sim \chi \bec{\nabla}^2 s$. In the latter estimate we assumed that: (i) the temperature fluctuations are much smaller than the mean fluid temperature; (ii) $\ell_0 \ll L_T$;
(iii) the coefficient of the molecular heat conductivity is independent of the coordinates;
(iv) for a low Mach numbers the entropy fluctuations are given by
\begin{eqnarray}
s= c_{\rm p} \left({\theta \over \meanT} + {(1-\gamma) p \over c_{\rm s}^2\meanrho}\right)
\approx c_{\rm p} {\theta \over \meanT},
\label{AC1}
\end{eqnarray}
where $c_{\rm s} = \left(\gamma\meanp/\meanrho\right)^{1/2}$ is the sound speed.
Equation~(\ref{AC1}) follows from the equation of state for a perfect gas: $P=(c_{\rm p}-c_{\rm v}) \rho T$ and the definition of entropy. Let us derive the equation for the
velocity-entropy correlation $\overline{{\bm u} s}$ in two limiting cases for small
and large P\'{e}clet number, where ${\rm Pe} = u_{0} \ell_0/\chi$ is the P\'{e}clet number
and $u_{0}$ is the characteristic turbulent velocity in the integral scale of turbulence,
$\ell_0$.

\subsection{Small P\'{e}clet numbers}

In order to study entropy fluctuations for small P\'{e}clet numbers we use a quasi-linear approach
\citep{M78,KR80}, that for a given velocity field is valid only for small P\'{e}clet numbers (${\rm Pe} \ll 1$).
In the framework of this approximation we neglect the nonlinear term and keep the molecular diffusion term in Eq.~(\ref{C1}). We rewrite Eq.~(\ref{C1}) in a Fourier space and solve this equation. The solution is:
\begin{eqnarray}
s(\omega, {\bm k}) = G_\chi(\omega, {\bm k}) I(\omega, {\bm k}),
\label{CC2}
\end{eqnarray}
where $G_\chi(\omega, {\bm k}) = (\chi k^2 + i \omega)^{-1}$, $\omega$ is the frequency, 
and ${\bm k}$ is the wave vector.
We apply a standard two-scale approach, whereby the non-instantaneous two-point second-order correlation function is written as follows:
\begin{eqnarray}
&& \overline{u_i(t_1, {\bm x}) \, s(t_2, {\bm  y})} = \int \overline{u_i (\omega_1, {\bm k}_1) s(\omega_2, {\bm k}_2)} \exp[i({\bm  k}_1 {\bm \cdot} {\bm x}
+ {\bm k}_2 {\bm \cdot} {\bm y})
\nonumber\\
&& \quad \quad + i(\omega_1 t_1 + \omega_2 t_2)] \,d\omega_1 \, d\omega_2 \,d{\bm k}_1 \, d{\bm k}_2 = \int F_i(\omega, {\bm k})  \exp[i {\bm k} {\bm \cdot} {\bm r} + i\omega \, \tilde \tau] \,d\omega \,d {\bm k} ,
\label{C2}
\end{eqnarray}
where we use large scale variables: ${\bm R} = ({\bm x} + {\bm y}) / 2$, $\, {\bm K} = {\bm k}_1 + {\bm k}_2$, $\, t = (t_1 + t_2) / 2$, $\, \Omega = \omega_1 + \omega_2$, and small scale  variables: ${\bm r} = {\bm x} - {\bm y}$, $\, {\bm k} = ({\bm k}_1 - {\bm k}_2) / 2$, $\, \tilde \tau = t_1 - t_2$, $\, \omega = (\omega_1 - \omega_2) / 2$, and
\begin{eqnarray}
&& F_i(\omega, {\bm k}) = \int \overline{u_i(\omega_1, {\bm k}_1) \, s(\omega_2, {\bm k}_2)} \exp[i \Omega t + i {\bm K} {\bm \cdot} {\bm R}] \,d \Omega \,d {\bm  K} .
\label{C3}
\end{eqnarray}
Here $\omega_1 = \omega + \Omega / 2$, $\, \omega_2 = - \omega + \Omega / 2$, ${\bm k}_1 = {\bm k} + {\bm  K} / 2$, and ${\bm k}_2 = - {\bm k} + {\bm  K} / 2$ (see, e.g., \cite{RS75}). We assume here that there is a separation of scales,
i.e., the maximum scale of random motions $\ell_0$ is much
smaller than the characteristic scales of inhomogeneities of the
mean entropy and fluid density.
Equations~(\ref{CC2})-(\ref{C3}) yield the velocity-entropy correlation $\overline{{\bm u} s}$:
\begin{eqnarray}
& & \overline{u_i s} = \int \overline{u_i(\omega, {\bm k}) \, I(- \omega, - {\bm k})} \, G_\chi^\ast \,d \omega \,d {\bm  k} = - (\nabla_j \means) \, \int \overline{u_i(\omega, {\bm k}) \, u_j(- \omega, - {\bm k})} \, G_\chi^\ast \,d \omega \,d {\bm  k}.
\nonumber\\
\label{C5}
\end{eqnarray}
We use the simple model for the second moments,
$\overline{u_i(\omega, {\bm k}) \, u_j(-\omega, -{\bm k})}$,
of a random velocity field in a Fourier space
for inhomogeneous, isotropic and incompressible flow:
\begin{eqnarray}
\overline{u_i(\omega, {\bm k}) \, u_j(-\omega, -{\bm k})} &=&  {\tilde E(k) \, \Phi(\omega) \over 8 \pi k^2} \Big[\delta_{ij} - {k_i \, k_j \over k^2} + {i \over 2 k^2} \, \big(k_i \nabla_j - k_j \nabla_i\big)\Big] \overline{{\bm u}^2}  .
\label{C4}
\end{eqnarray}
This model is obtained using the symmetry arguments \citep{BA71}.
Here $\delta_{ij}$ is the Kronecker tensor, the energy spectrum function
is $\tilde E(k) = C_E \, k_0^{-1} \, (q-1) \, (k / k_{0})^{-q}$ for the
range of wavenumbers $k_0<k<k_d$, the wavenumber $k_{0} = 1 / \ell_0$, the length $\ell_0$ is the maximum scale of random motions, the exponent $1<q<3$, and the coefficient $C_E=[1 - (k_0/k_d)^{q-1}]^{-1}$.
We use the frequency function $\Phi(\omega)$ in the form of the
Lorentz profile: $\Phi(\omega)=[\pi \tau_c \,
(\omega^2 + \tau_c^{-2})]^{-1}$, where $\tau_c$ is the correlation time of
a random velocity field. This model for
the frequency function corresponds to the following non-instantaneous correlation function:
$\overline{u_i(t) u_j(t+\tau)} \propto \exp (-\tau / \tau_c)$.

We use Eqs.~(\ref{C5}) and~(\ref{C4}), and after integration in $\omega$-space and in ${\bm k}$-space in Eq.~(\ref{C5}) we obtain the formula for the velocity-entropy correlation $\overline{u_i \, s}$:
\begin{eqnarray}
\overline{u_i \, s} &=& - \chi_T \nabla_i \means \;,
\label{C6}\\
\chi_T &=& C_\chi \, u_{0} \, \ell_0 \, {\rm Pe}, \quad C_\chi = {q-1\over 3(q+1)} \, \left[{1 - (k_0/k_d)^{q+1} \over 1 - (k_0/k_d)^{q-1}}
\right] ,
\label{C7}
\end{eqnarray}
where $\chi_T$ is the coefficient of the turbulent diffusion of the mean entropy
and $u_{0}=\sqrt{\overline{{\bm u}^2}}$  is the characteristic velocity in the maximum scale of random motions.
Here we used that $I_0 \equiv \int \Phi(\omega) G_\chi(\omega, {\bm k}) \, d\omega = \tau_c /(1+\tau_c \eta k^2)$, and for small P\'{e}clet numbers $I_0 \approx (\eta k^2)^{-1}$.
The coefficient $C_\chi=1/3$ for a narrow range of the random velocity field
in the wavenumbers, $k_d-k_0 \ll k_d$, and $C_\chi=(q-1)/3(q+1)$ for a wide range in the wavenumbers, $k_d \gg k_0$.
Contributions (which are proportional to $\nab$ in Eq.~(\ref{C4})) to the velocity-entropy correlation $\overline{u_i \, s}$, after the integration over the angles in ${\bm k}$-space, vanish. However, the coefficient of the turbulent diffusion $\chi_T$ depends on the coordinates, due to the inhomogeneous turbulence.
Equations~(\ref{C6}) and~(\ref{C7}) are in agreement with those obtained by means of dimensional arguments \citep{BH59} and by the Lagrangian-history direct-interaction approximation \citep{KR68}.

\subsection{Large P\'{e}clet numbers}

In this subsection we derive a formula for the velocity-entropy correlation $\overline{{\bm u} s}$ using the spectral $\tau$ approach that is valid for large P\'{e}clet and Reynolds numbers
(${\rm Pe} \gg 1$). Using Eq.~(\ref{C1}) written in a Fourier space we derive equation for the instantaneous two-point second-order correlation functions $F_i(t, {\bm k}) = \overline{u_i(t, {\bm k}) \, s(t, -{\bm k})}$:
\begin{eqnarray}
{dF_i \over dt} &=& \overline{u_i(t, {\bm k}) \, I(t, -{\bm k})} + \hat{\cal M} F_i^{(III)}({\bm k}) \,,
\label{BD1}
\end{eqnarray}
where $\hat{\cal M} F_i^{(III)}({\bm k}) = - [\overline{u_i \, {\cal N}} + \overline{(\partial u_i / \partial t) \, s} - \chi \overline{u_i \, \bec{\nabla}^2 s}]_{\bm k}$ are the third-order moment terms appearing due to the nonlinear terms which also include the molecular diffusion term.

The equation for the second moment includes the first-order spatial
differential operators applied to the third-order
moments. A problem arises regarding how to close the system, i.e.,
how to express the third-order terms $\hat{\cal M}
F^{(III)}$ through the lower moments $F^{(II)}$
\citep{O70,MY75,Mc90}. We use the spectral $\tau$ approximation which postulates that the deviations of the third-moment terms, $\hat{\cal M} F^{(III)}({\bm k})$, from the contributions to these terms by the background turbulence, $\hat{\cal M} F^{(III,0)}({\bm k})$, can be expressed through similar deviations of the second moments, $F^{(II)}({\bm k}) - F^{(II,0)}({\bm k})$ \citep{O70,PFL76}:
\begin{eqnarray}
&& \hat{\cal M} F^{(III)}({\bm k}) - \hat{\cal M} F^{(III,0)}({\bm
k})= - {1 \over \tau_r(k)} \, \Big[F^{(II)}({\bm k}) - F^{(II,0)}({\bm k})\Big] \,,
\label{D2}
\end{eqnarray}
where $\tau_r(k)$ is the scale-dependent relaxation time, which can be identified with the correlation time $\tau(k)$ of the turbulent velocity field for large Reynolds and P\'{e}clet numbers. The functions with the superscript $(0)$ correspond to the background turbulence with a zero gradient of the mean entropy. Validation of the $\tau$ approximation for different situations has been performed in numerous numerical simulations and analytical studies (see e.g. the review by \cite{BS05}; and also discussions by \cite{RK07,RKKB11}).

Note that the contributions of the terms with the superscript $(0)$ vanish because when the gradient of  the mean entropy is zero, the turbulent heat flux and the entropy fluctuations vanish. Consequently, Eq.~(\ref{D2}) for $\hat{\cal M} F_i^{(III)}({\bm k})$ is reduced to $\hat{\cal M} F_i^{(III)}({\bm k}) = - F_i({\bm k}) / \tau(k)$.
We also assume that the characteristic time of variation of the second moment $F_i({\bm k})$ is substantially larger than the correlation time $\tau(k)$ for all turbulence scales. Therefore, in a steady-state Eq.~(\ref{BD1}) yields the following formula for the velocity-entropy correlation:
\begin{eqnarray}
F_i &=& \int \tau(k) \, \langle u_i(t, {\bm k}) \, I(t, -{\bm k}) \rangle \, d{\bm k} =
- (\nabla_j \means) \, \int \tau(k) \, \langle u_i({\bm k}) \, u_j(-{\bm k}) \rangle \, d{\bm k} .
\label{BD3}
\end{eqnarray}
We use the following simple model for the second moments,
$\overline{u_i({\bm k}) \, u_j(-{\bm k})}$,
of a turbulent velocity field in Fourier space
for inhomogeneous, isotropic and incompressible flow for large Reynolds numbers:
\begin{eqnarray}
\overline{u_i({\bm k}) \, u_j(-{\bm k})} = {\tilde E(k) \over 8 \pi k^2} \Big[\delta_{ij} - {k_i \, k_j \over k^2} + {i \over 2 k^2} \, \big(k_i \nabla_j - k_j \nabla_i\big)\Big] \overline{{\bm u}^2}.
\label{D3}
\end{eqnarray}
This model is obtained using symmetry arguments \citep{BA71}.
After integration in ${\bm k}$-space of Eq.~(\ref{BD3}) we arrive at an equation for the velocity-entropy correlation, $\langle u_i \, s \rangle$:
\begin{eqnarray}
\langle u_i \, s \rangle &=& - \chi_T \nabla_i \means \;, \quad \chi_T = u_{0} \, \ell_0 /3 ,
\label{LC6}
\end{eqnarray}
where $u_{0}=\sqrt{\overline{{\bm u}^2}}$ is the characteristic turbulent velocity.
In the derivation of Eq.~(\ref{LC6}) we used the following expression for the turbulent correlation time: $\tau(k) = 2 \, \tau_0 \, (k / k_{0})^{1-q}$, where $\tau_0 = \ell_0 / u_{0}$ is the characteristic turbulent time.
Contributions (which are proportional to $\nab$ in Eq.~(\ref{D3})) to the the velocity-entropy correlation $\overline{u_i \, s}$, after the integration over the angles in ${\bm k}$-space,  vanish. However, the coefficient of turbulent diffusion $\chi_T$ depends on the coordinates, due to the inhomogeneous turbulence.
Therefore, the formulae for the velocity-entropy correlation, $\langle u_i \, s \rangle$, are similar for small and large P\'{e}clet numbers, while the coefficients of turbulent diffusion of the mean entropy are different in these two limiting cases. Equation~(\ref{LC6}) is in agreement with that derived by means of the path integral approach \citep{EKR95}, by dimensional arguments and by the renormalization procedure used for large P\'{e}clet numbers \citep{EKRS96}.

\section{Conclusions}

In the present study we have demonstrated that
for a low-Mach-number compressible fluid flow the turbulent flux of entropy,
${\bm F}_s=\meanrho \, \overline{{\bm u} s}=-\meanrho \, \chi_T \nab \means$,
is different from the turbulent convective flux of the mean internal energy,
${\bm F}_c=\meanT \,\meanrho \, \overline{{\bm u} s}
=-\meanT \, \meanrho \, \chi_T \nab \means$.
As follows from the analysis performed in Section 3,
the coefficient of turbulent
diffusion of entropy $\chi_{\rm T}$ depends on the P\'{e}clet number.
For small P\'{e}clet numbers, applying the quasi-linear approach for an isotropic and inhomogeneous
background random velocity field we obtain the following coefficient of turbulent diffusion of the mean entropy: $\chi_T= C_\chi {\rm Pe} \, u_{0} \ell_0$,
where the constant $C_\chi$ depends on the energy spectrum
of the random velocity field.
For large P\'{e}clet and Reynolds numbers,
applying the  spectral $\tau$ approximation we get the following coefficient of turbulent diffusion of the mean entropy: $\chi_T = u_{0} \ell_0 /3$.

\bigskip

\begin{acknowledgements}
We are indebted to A. Brandenburg, E. Dormy, P. J. K\"apyl\"a,
M. Rheinhardt and P.~H.~Roberts for stimulating discussions.
This work was supported in part by
the Research Council of Norway under the FRINATEK (grant No. 231444) and
the Academy of Finland (grant No. 280700).
\end{acknowledgements}

\newpage
\bibliographystyle{jpp}

\bibliography{paper}

\end{document}